\documentclass[sigconf, 9pt]{acmart}

\renewcommand\footnotetextcopyrightpermission[1]{}

\AtBeginDocument{%
  }

\usepackage{amsmath}
\usepackage{algorithmic}
\usepackage{graphicx}
\usepackage{textcomp}
\usepackage{xcolor}
\usepackage{booktabs} 
\usepackage{soul}
\graphicspath{{figs/}}
\usepackage{braket}
\usepackage{etoolbox}
\usepackage{fancyhdr}

\setcopyright{acmcopyright}
\copyrightyear{2018}
\acmYear{2018}
\acmDOI{XXXXXXX.XXXXXXX}

\acmConference[Conference acronym 'XX]{Make sure to enter the correct
  conference title from your rights confirmation emai}{June 03--05,
  2018}{Woodstock, NY}
\acmPrice{15.00}
\acmISBN{978-1-4503-XXXX-X/18/06}

\begin{document}

\fancyhead{}
\title{\vspace{-0.5em}Evaluating Efficacy of Model Stealing Attacks and Defenses on Quantum Neural Networks}

\author{Satwik Kundu}
\affiliation{%
  \institution{The Pennsylvania State University}
  \city{State College}
  \state{Pennsylvania}
  \country{USA}
  \postcode{16801}}
\email{sxk6259@psu.edu}

\author{Debarshi Kundu}
\affiliation{%
  \institution{The Pennsylvania State University}
  \city{State College}
  \state{Pennsylvania}
  \country{USA}
  \postcode{16801}}
\email{dqk5620@psu.edu}

\author{Swaroop Ghosh}
\affiliation{%
  \institution{The Pennsylvania State University}
  \city{State College}
  \state{Pennsylvania}
  \country{USA}
  \postcode{16801}}
\email{szg212@psu.edu}

\renewcommand{\shortauthors}{Kundu et al.}

\begin{abstract}

Cloud hosting of quantum machine learning (QML) models exposes them to a range of vulnerabilities, the most significant of which is the model stealing attack. 
In this study, we assess the efficacy of such attacks in the realm of quantum computing. 
We conducted comprehensive experiments on various datasets with multiple QML model architectures. Our findings revealed that model stealing attacks can produce clone models achieving up to $0.9\times$ and $0.99\times$ clone test accuracy when trained using Top-$1$ and Top-$k$ labels, respectively ($k:$ num\_classes). To defend against these attacks, we 
leverage the unique properties of current noisy hardware and perturb the victim model outputs and hinder the attacker's training process. In particular, we propose: 1) hardware variation-induced perturbation (HVIP) and 2) hardware and architecture variation-induced perturbation (HAVIP). Although noise and architectural variability can provide up to $\sim16\%$ output obfuscation, our comprehensive analysis revealed that models cloned under noisy conditions tend to be resilient, suffering little to no performance degradation due to such obfuscations. Despite limited success with our defense techniques,  this outcome has led to an important discovery: QML models trained on noisy hardwares are naturally resistant to perturbation or obfuscation-based defenses or attacks.
\end{abstract}




\maketitle

\section{Introduction}
Quantum machine learning (QML) merges quantum computing with classical machine learning, opening new horizons in computational speed and capability \cite{
schuld2015introduction}. Quantum neural networks (QNNs) are a notable development in QML, mirroring the structure and function of traditional neural networks within a quantum framework \cite{
abbas2021power}. While deep learning technologies, including DNNs, GNNs, and transformers, are transforming data processing, quantum computing is concurrently stepping into its long-envisioned potential. Recent advancements in parameterized quantum circuits (PQC) have demonstrated their potential in tackling previously unsolvable combinatorial optimization and molecular energy challenges \cite{
nakagawa2023analytical, hao2022exploiting}. These breakthroughs are catalyzing the growth of quantum deep learning, building on foundational deep learning research. PQCs, fundamental to QNNs, have been applied in diverse areas, from quantum chemistry simulations and finance optimization \cite{herman2022survey} to early disease detection \cite{kumar2021heart}. As applications broadens, the call for refined quantum models becomes more pressing.

Training QML models is a complex task, intertwining quantum computing's nuances with conventional machine learning. Key challenges involve efficient data encoding, managing qubit constraints for feature representation, and appropriately determining crucial hyperparameters, such as learning rate and QNN structure. 
Noise in today's NISQ-era quantum devices further complicates QNN training. Given the nascent state of quantum technologies, training a QML model demands significant resources and multiple optimization iterations. The rising necessity for large quantum datasets, coupled with expert knowledge and computational power, underscores the promise of Quantum Machine-Learning-as-a-Service (QMLaaS) \cite{saiwa}. As quantum infrastructure remains intricate and costly, many entities are poised to adopt QMLaaS, enabling them to leverage QNN benefits without direct infrastructural challenges.

\subsection{Motivation} 
\noindent \textbf{QMLaaS Vulnerabilites: }Training QML models requires considerable time, complexity, and quantum resources. To safeguard their intellectual property, soon model owner will provide access only through input-output queries via specific APIs, leading to the rise of QMLaaS. A similar approach, highlighted in classical machine learning, offers users sophisticated models on a pay-per-query basis, eliminating the challenges of data collection, optimization, and training. While QMLaaS would democratize access to cutting-edge PQC based models, it also exposes them to adversarial attacks on cloud platforms. Notably, there's a growing concern over ``model stealing" attacks, where attackers can replicate a model's architecture and behavior by systematic querying \cite{oliynyk2023know}.
Attackers may be motivated to steal models to (a) bypass restrictions on future PQC models available only through black-box APIs, (b) avoid paying fees for the services,  (c) create synthetic datasets in fields like drug discovery and finance where original datasets are costly using GANs \cite{kariyappa2021maze}, and (d) transition from black-box to white-box attack strategies by accessing the model's internals \cite{
sethi2018data}. This would not only jeopardizes the intellectual property rights of these proprietary PQC-based models but also casts shadows on the security robustness of future QMLaaS platforms.

Drawing from the vulnerabilities, it becomes imperative to probe the potential security challenges facing QMLaaS \cite{saiwa}. Given that classical models, despite their matured security protocols, can be breached, it raises concerns that QNNs, still in their formative stages, might be even more exposed. As the quantum computing landscape matures and QNNs gain traction, comprehending and countering the risks of model stealing is of utmost importance. This paper analyzes these potential vulnerabilities to fortify the next wave of QML models.

\noindent \textbf{Need to Study Model Stealing Attack on QNN: }
QNNs differ fundamentally from DNNs due to their reliance on PQCs as their core architectural component. Entanglement in PQCs facilitates more complex data manipulation, and the increased effective dimension enables them to represent more complex data. PQCs are also inherently sensitive to noise and errors due to fragile nature of quantum states in contrast to DNNs, which are generally robust to errors. The variability in noise across different quantum hardware further complicates the predictability of QNN behavior, especially in the context of classical model stealing attacks and defenses. The above aspects warrant a fresh look at QNN model stealing attacks. 


\subsection{Contribution}
In this study, 1) we assess the effectiveness of model stealing attacks on hybrid QNNs across diverse datasets \emph{for the first time} to the best of our knowledge. When viewing the cloud-based QNN as a black-box, an attacker can iteratively query the victim QNN for inputs, subsequently building a dataset from the returned outputs. Using this acquired dataset, the adversary can then train a clone model that mirrors the performance of the original, stealing it's functionality, 2) capitalize on the inherent noise and architectural variability to propose two perturbation-based defense methods: (a) hardware variation-induced perturbation (HVIP) and (b) hardware and architecture variation-induced perturbation (HAVIP) and, 3) show that training QNNs in noisy conditions actually increases their resistance to perturbation-based defenses, implying a heightened robustness against perturbation-based adversarial attacks as well.


\begin{figure}[!t]
        \vspace{0mm}
        \centering 
        \includegraphics[width=0.95\linewidth]{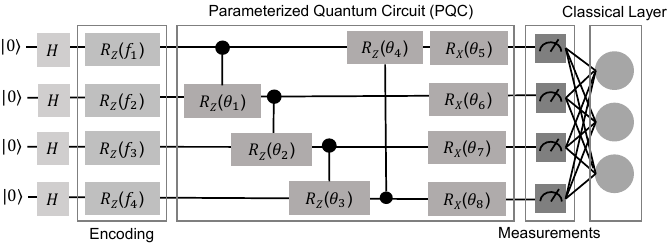}
        \vspace{-3mm}
        \caption{Architecture of a 4-qubit hybrid QNN. Classical features are encoded as angles of quantum rotation gates ($R_Z$). PQC transforms encoded states to explore the search space and entangle features. Measured expectation values are then fed into a classical linear layer for final prediction.}
        \label{qnn_circuit}
        \vspace{-4mm}
\end{figure}

\section{Background} \label{background}

\subsection{Quantum Neural Network (QNN)} 
QNN mainly consists of three building blocks: (i) a classical to quantum data encoding (or embedding) circuit, (ii) a parameterized quantum circuit (PQC) whose parameters can be tuned (mostly by an optimizer) to perform the desired task, and (iii) measurement operations. There are a number of different encoding techniques available (basis encoding, amplitude encoding, etc.) but for continuous variables, the most widely used encoding scheme is angle encoding where a variable input classical feature is encoded as a rotation of a qubit along the desired axis \cite{abbas2021power}. As the states produced by a qubit rotation along any axis will repeat in 2$\pi$ intervals, features are generally scaled within 0 to 2$\pi$ (or -$\pi$ to $\pi$) in a data pre-processing step. In this study, we used $RZ$ gates to encode classical features into their quantum states.

A PQC consists of a sequence of quantum gates whose parameters can be varied to solve a given problem. In QNN, the PQC is the primary and only trainable block to recognize patterns in data. The PQC is composed of entangling operations and parameterized single-qubit rotations. The entanglement operations are a set of multi-qubit operations (may or may not be parameterized) performed between all of the qubits to generate correlated states and the parametric single-qubit operations are used to search the solution space. 
Finally, the measurement operation causes the qubit state to collapse to either `0' or `1'. We used the expectation value of Pauli-Z to determine the average state of the qubits. The measured values are then fed into a classical neuron layer (the number of neurons is equal to the number of classes in the dataset) in our hybrid QNN architecture as shown in Fig. \ref{qnn_circuit}, which performs the final classification task.

\begin{figure*}[!t]
    \vspace{-2mm}
    \centering
    \includegraphics[width=0.9\textwidth]{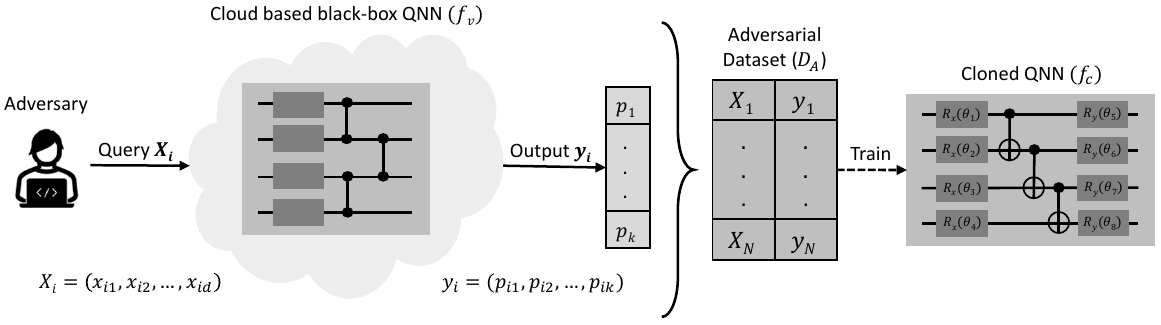}
    \vspace{-2mm}
    \caption{Adversary sends a query $X_i = (x_{i1}, x_{i2}, ..., x_{id})$, where $d$ refers to dimension of input vector, to the cloud based victim QNN which is represented using $f_v$. The QNN returns a vector of class probabilities as output i.e., $f_v(X_i) = y_i = (p_{i1}, p_{i2}, ..., p_{ik})$ where $k$ is the number of classes of the dataset on which original cloud-based model is trained. Adversary repeats this process multiple times to create the attacker dataset $D_A$. The attacker then trains a substitute model $f_c$ to clone the functionality of $f_v$.} 
    \label{attack}
    \vspace{-4mm}
\end{figure*}

\subsection{Classical Model Stealing Attacks and Defenses}
\noindent \textbf{Overview:} In model stealing attacks, adversaries target a ``victim/target model", treating it as an ``oracle" to gather input-output pairs \( (X,y) \). Each interaction with the model, termed a "query", provides insights into its operation. When only the model's outputs are accessible, it's considered a "black-box" attack. Conversely, if an attacker knows the model's complete details, the attack is "white-box". Any middle ground is labeled as "grey-box". 
Broadly, model stealing attacks can be segmented into two primary objectives:



\begin{itemize}
    \item \textit{Stealing Exact Model Properties:} Hyperparameter attacks aim to uncover parameters like regularization used during model training. 
    Architecture attacks focus on deducing the neural network's design, such as layer count, layer types, etc. 
    Learned parameters attacks aim to extract specific weights and parameters from a known architecture, enabling attackers to replicate the target model's behavior.
    \item \textit{Stealing Model Behavior:} Stealing model behavior encompasses two primary goals: (a) achieve same level of effectiveness where the attacker creates a model equivalent in performance to the target, potentially using a similar or completely different architecture and (b) ensure the replicated model's predictions consistently align with the target's, even in misclassifications. 
\end{itemize}

\noindent \textbf{Attack approaches:} Various model stealing attack approaches exist, including Witness-finding Attack \cite{tramer2016stealing, reith2019efficiently}, 
Equation-solving attacks \cite{yan2021monitoring}, Path-finding attacks \cite{tramer2016stealing}, Recovering attacks\cite{rolnick2020reverse}, and Substitute model training. Among these, substitute model training stands out as the most widely adopted technique \cite{oliynyk2023know}. This method involves training a model on data labeled by the target, effectively replicating its functionality.

\noindent \textbf{Defense techniques:}
Attack prevention strategies aim to mitigate the effectiveness of model stealing, rather than blocking the theft itself. The goal is to degrade the quality of the stolen model to render it useless. Most widely used technique for protecting models against these types of attacks is data perturbation, where either the model's input or output is altered to maintain model integrity while producing less precise results. Basic defenses limit the information returned by models to mere labels \cite{tramer2016stealing}. Authors in \cite{wang2020information} introduced concept of Information Laundering (IL), simultaneously perturbs input and output to preserve model utility and confidentiality. 

\section{Proposed Attack and Defense Methodologies}
\subsection{Adversary's Knowledge}
Here, we assume QNN model hosted on the cloud, treated as a black box. This means the user, acting as an adversary, is unaware of the model's architecture, size (i.e., number of qubits and number of parametric layers), hyperparameters, or the dataset on which it was originally trained. The only knowledge the adversary possesses is the input-output structure; they understand the format of the data the QNN model accepts and how it presents its output. Furthermore, the adversary remains uninformed about the quantum device(s) on which the circuit was trained or on which device the circuit executes during the inference phase—when they query the model. Using only this limited input-output data, the adversary attempts to construct an adversarial dataset $D_A$. Their aim is to train a cloned model, leveraging this dataset, which can mimic the cloud-hosted QNN's functionality.

\subsection{Attack Strategy}
In Fig. \ref{attack}, we illustrate the complete attack procedure which resembles classical model stealing techniques. Given input, $X_i$, consisting of features $(x_{i1}, x_{i2}, ..., x_{id})$, where $d$ represents the number of features per sample which the victim QNN takes as input, an adversary sends queries to a cloud-based, trained QNN. The victim QNN, represented as \(f_v(X_i)\), executes the consequent circuit on a quantum device. The quantum device returns measured expectation values for the qubits as output. Subsequently, these values are fed into a classical linear layer. Notably, the number of neurons in this layer equals the total class count. This layer then produces the softmax probability vector, \(y\), for each class. The combination of \(X_i\) and \(y_i\) allows the adversary to generate an adversarial dataset that mirrors the functionality of the victim model, \(f_v\). Once in possession of this dataset, the adversary can train a cloned QNN model, \(f_c\). The intent is to make this clone, \(f_c\), operate in a manner closely resembling the victim QNN, symbolized as \(f_v \sim f_c\). 

The challenge at hand involves identifying the optimal method to create an unlabeled dataset. This dataset will be used to probe a victim QNN, which will subsequently help in constructing the attacker's dataset, denoted as \(D_A\). The ultimate objective is to craft a \(D_A\) that, when used to train another model, mimics the performance of the original QNN closely. This becomes especially intricate when the attacker is oblivious to the data that trained the original model. If successful, a sufficiently diverse \(D_A\) may not only achieve a similar performance to the original cloud-based model but could also become a valuable asset that the attacker could potentially sell. This is because any model trained on such a dataset might reflect the functionality of the victim QNN. Previous approaches to this challenge have varied, with some merging problem domain (PD) with non-problem domain (NPD) data. Others have even experimented with generating synthetic data, by sampling from a specific probability distribution or by incorporating noise, etc. We assume the attacker lacks access to the original training dataset. Therefore, we perform experiments by; 1) mixing NPD datasets, and 2) creating a random synthetic dataset, which we use as the unlabeled dataset to query the victim QNN.

\subsection{Defense Strategies}
\begin{figure}[!t]
        \vspace{0mm}
        \centering 
        \includegraphics[width=\linewidth]{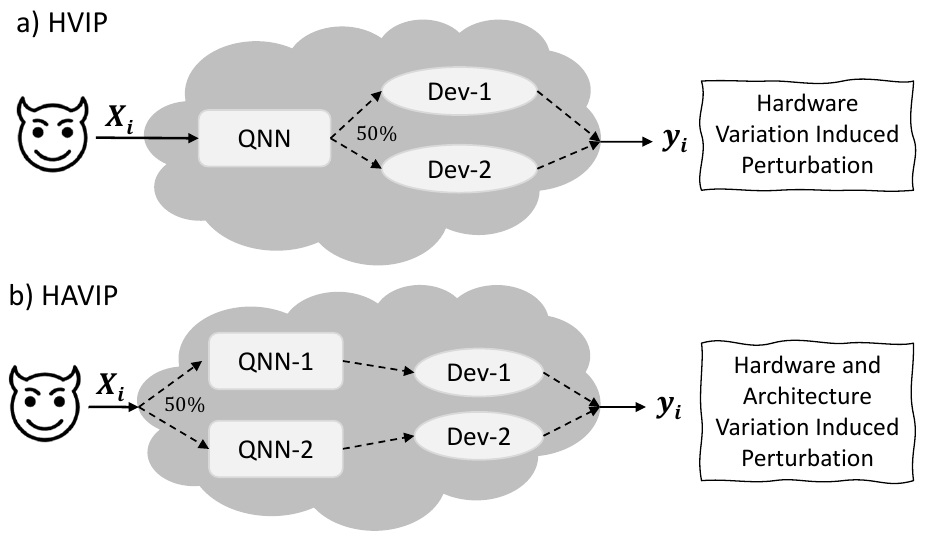}
        \vspace{-4mm}
        \caption{Proposed defense techniques; a) HVIP: Victim randomly (with 50\% probability) alternates between quantum devices when executing the QNN to perturb the output. b) HAVIP: Victim has trained multiple QNNs on multiple devices which the users are unaware of. During query stage, the victim randomly sends query to any QNN for inference which should help in further obfuscation of output values.}
        \label{defense}
        \vspace{-4mm}
\end{figure}

Model stealing attacks are notoriously difficult to defend against \cite{oliynyk2023know, kariyappa2020defending}. Many defense techniques attempt to introduce unique strategies that perturb the output of the victim model. By doing so, attackers find it challenging to train their clone model effectively due to the noisy dataset obtained from the target model. In the realm of quantum computing, NISQ (Noisy Intermediate-Scale Quantum) devices already contain inherent noises. The victim can attempt to leverage these inherent noises as a means to autonomously disrupt output values, thereby safeguarding their cloud-hosted QNNs. We capitalize on the device-to-device variability of NISQ machines to defend against model stealing attacks. Specifically, we introduce two main defense techniques: a) Hardware Variation-Induced Perturbation (HVIP), and b) Hardware and Architecture Variation-Induced Perturbation (HAVIP) (Fig. \ref{defense}).

\noindent \textbf{a) HVIP:} 
Rather than consistently executing the quantum circuit on the same device for each attacker query, the victim can dynamically alternate between various quantum devices that differ in terms of coupling maps, basis gates, readout errors, and gate errors naturally perturbing the output vectors. These inherent variations can obscure the measurement output, thereby making it harder for the attacker to train a clone model that effectively mirrors the functionality of the cloud-based QNN. In classical model stealing techniques, this isn't feasible because the hardware used for training, such as GPUs are ideal i.e., they lack inherent noise. Consequently, executing the model on different GPUs during the inference stage would yield identical outputs.

\noindent \textbf{b) HAVIP:} To further improve robustness of cloud-based QNN against these kind of model stealing attacks, the victim can secretly train various QNN models on different devices and then host them on the cloud. This concealed setup means attackers remain unaware of the total number and types of QNN models available. For each incoming query, the victim randomly selects a QNN and executes it on a particular quantum hardware, delivering the resulting outputs. This strategy, known as HAVIP, presents a more formidable obfuscation compared to HVIP. The essence of its strength lies in the variety: different QNN architectures inherently perform differently, yielding diverse probability distribution vectors. This variance is amplified when executed on different devices; post-transpilation, the compiled circuits can vary drastically in gate counts, depths, and consequent errors—attributes like the number of swap gates, qubit quality, and total gate error can differ significantly which would further obfuscate the outputs. Consider a scenario where the victim employs two unique QNN architectures, each trained on a distinct quantum device. A random inference from any of them could provide the attacker with a noisy dataset, making it hard to accurately train the clone model.

\section{Evaluation}
\subsection{Setup} \label{setup}
\noindent \textbf{Training:} For evaluation, we used multiple different circuits from \cite{sim2019expressibility} to build our victim and clone model (PQC-$x$ represents circuit-$x$ in \cite{sim2019expressibility}), initialized with random weights. In all experiments we used a 4-qubit QNN as our victim and clone models (if not mentioned otherwise). For embedding classical features to their corresponding quantum state we use the angle encoding technique with $RZ$ gate and for measurement, we calculate the Pauli-Z basis expectation value over all the qubits. For training QNN when only the top label is available we used the NLLLoss() and used KLDivergence() when the full probability vectors is available. The hyperparameters used for training both victim and clone model are; Epochs: 25, learning\_rate ($\eta$): 0.01, batch\_size = 32 and optimizer: Adam. All training are done on an Intel Core-i7-12700H CPU with 40GB of RAM.

\begin{table}[!b]
    \vspace{-2mm}
    \centering
    \caption{The final test accuracy and loss for both the victim and clone QNN after 25 epochs of noisy training on various datasets. It is evident that training with probability vectors (top-$k$, $k$: num\_classes) yields significantly better performance than using only the predicted label (Top-$1$).}
    \label{acc_loss}
    \vspace{-2mm}
    \begin{tabular}{ccccccc}
    \cmidrule(lr){2-7}
    \multicolumn{1}{c}{} & \multicolumn{2}{c}{Victim} & \multicolumn{2}{c}{Clone (Top-$1$)} & \multicolumn{2}{c}{Clone (Top-$k$)} \\
    \cmidrule(lr){1-7}
    Datasets    & Acc. & Loss & Acc. & Loss & Acc. & Loss \\
    \cmidrule(lr){1-7}
    MNIST-4     & 0.896      & 0.371      & 0.726     & 0.691     & 0.880      & 0.547      \\
    Fashion-4     & 0.856      & 0.475      & 0.776      & 0.625     & 0.823      & 0.590        \\ 
    Kuzushiji-4   & 0.796      & 0.709      & 0.680     & 0.782     & 0.776      & 0.749       \\  
    \bottomrule
    \end{tabular}
    \vspace{-1mm}
\end{table}

\begin{figure}[!b]
    \vspace{-1mm}
    \centering
    \includegraphics[width=\linewidth]{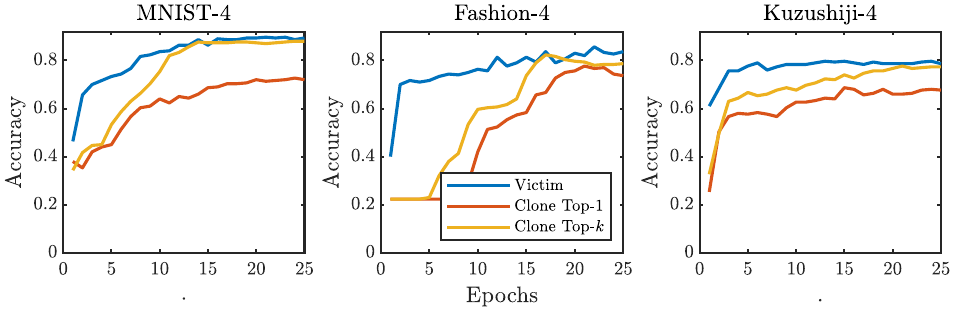}
    \vspace{-6mm}
    \caption{Comparison of test accuracy between the victim and the clone model trained with Top-$1$ and Top-$k$ labels across various datasets. The clone model, when trained using the Top-$k$ vector, closely mirrors the performance of the victim due to the richer information per training sample.} 
    \label{acc_loss_fig}
    \vspace{0mm}
\end{figure}

\begin{figure*}[!t]
    \vspace{-4mm}
    \centering
    \includegraphics[width=0.8\textwidth]{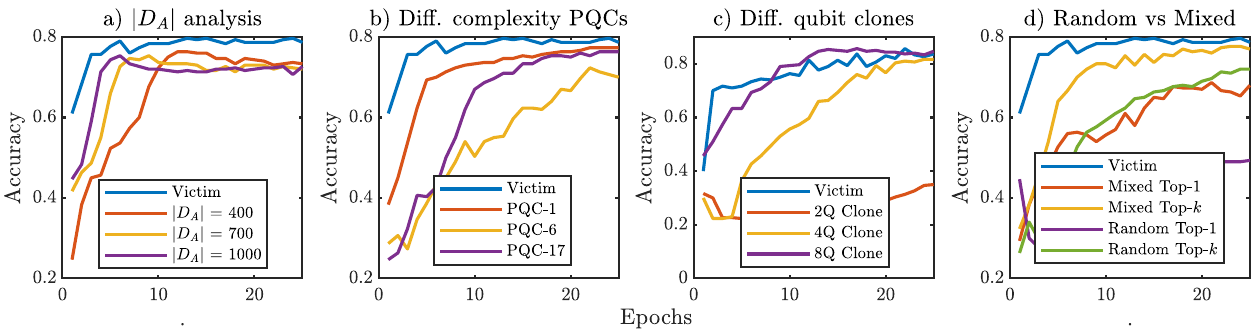}
    \vspace{-3mm}
    \caption{Plots comparing test accuracies for clone models; a) trained using different sized datasets ($|D_A|$) , b) with different sized PQCs i.e. different circuit depth and gate count but same qubit count, c) with different width QNNs i.e. different qubit clone models and d) when trained using mixed i.e., merging NPD datasets vs random dataset.} 
    \label{attack_analysis}
    \vspace{-4mm}
\end{figure*}

\noindent \textbf{Device:} We performed all experiments on Pennylane's \cite{bergholm2018pennylane} ``default.mixed" device. Due to the long queue times inherent to real quantum hardware, and in an effort to mimic the performance on these noisy devices, we introduced custom noises such as readout\_error, amplitude damping, phase flip, depolarizing error and BitFlip error to the ``default.mixed" device. Additionally, to accurately assess the post-compilation performance of circuits, mirroring their execution on actual quantum hardware, we incorporated custom basis gates (similar to IBMQ devices). Given that classical gradient calculation techniques like backpropagation are not applicable on real quantum devices, all our QNN training employed the practically feasible SPSA gradient calculation method. This produces noisy gradients, ensuring our results aligns with those from real hardware.

\noindent \textbf{Dataset:} We conduct all attack and defense experiments using a reduced feature set of MNIST, Fashion, Kuzushiji and Letters datasets with latent dimension $d = 8$ (from the original 28$\times$28 sized image) generated using a convolutional autoencoder \cite{alam2021quantum}. Thus, for each dataset, we create a smaller 4-class dataset from these reduced feature sets i.e., MNIST-4 (class 0, 1, 2, 3), Fashion-4 (class 6, 7, 8, 9), Kuzushiji-4 (class 3, 5, 6, 9) and Letters-4 (class 1, 2, 3, 4) with each having 1000 samples (700 for training and 300 for testing). Since we use 4-qubit QNN models for training and each of these datasets is of dimension $d = 8$, we encode 2 features per qubit. Number of shots/trials is set to 1000 for all experiments.

For creating the unlabeled dataset which attacker uses for querying the victim QNN, we use a mix of non-problem domain (NPD) datasets. For example, if the victim QNN is trained on MNIST-4, we use an unlabeled dataset formed by mixing features of Fashion-4, Kuzushiji-4 and Letters-4 and similarly for the other cases. Since, the original model is trained on 700 samples, we used datasets of size exactly 700 to query and train our victim model. However, we experimented with random datasets as well as different sized attacker dataset ($D_A$) with results presented in subsequent sections.

\subsection{Results and Analysis of Attack}  
\noindent \textbf{Top-1 vs. Top-$\boldsymbol{k}$:} Since attacker may be unaware of the device that was used to train the original model, we trained both victim and clone model on different devices with different noise characteristics to better replicate real world scenario. We evaluated the efficacy of model stealing attacks on hybrid QNNs under two conditions. Cloud-based QNN returns: 1) only the highest probability (Top-$1$) or 2) the complete probability vector for all classes (Top-$k$). The results from Table \ref{acc_loss} and Fig. \ref{acc_loss_fig} distinctly show that the cloned model, when trained with Top-$k$ labels, aligns more closely with the performance of the victim model than when trained using solely the Top-$1$ label. This can be attributed to the richer information offered by the full probability vectors. Such comprehensive data equips the attacker with a more descriptive dataset, allowing the cloned model to mirror the cloud-based QNN's more accurately. On an average, cloning using Top-$1$ and Top-$k$ labels resulted in clone test accuracy of 0.85$\times$ and 0.98$\times$ respectively.

\noindent \textbf{$\boldsymbol{|D_A|}$ Analysis:} We also experimented with different sized $D_A$ in-order to see how big of an impact does the size of $D_A$ (represented as $|D_A|$) have on the performance of clone model. This is particularly important since size of $|D_A|$ is directly proportional to the total cost incurred by the attacker to query and create $D_A$. If smaller sized dataset can be used, that would allow for more efficient model stealing. From Fig. \ref{attack_analysis}a) we can clearly see that eventhough larger dataset allows faster learning, the final test accuracy is similar for all cases. Thus, it reflects that the attacker is better off using smaller sized dataset compared to one used for training since larger dataset does not necessarily improve the performance of the clone model.

\noindent \textbf{Different PQCs:} Next, we compared the clone performance of models with different architectures. PQC-1 here is the least complex model with no entangling gates, followed by PQC-17 and PQC-6 being the most complex circuit in terms of circuit depth, gate count and entanglement \cite{sim2019expressibility}. The results from Fig. \ref{attack_analysis}b) show that the simpler circuit perform better than the more complex PQC-6 which indicates that simpler quantum circuits are better for model stealing. One possible explanation for the same might be that simpler model has lesser depth and gate counts thus the total error accumulated, especially after compilation is lower than the complex models.

\noindent \textbf{Architecture Width:} Given that the attacker lacks knowledge about the architecture of the cloud-based QNN, they are also unaware about the qubit count of the target model. Thus, we ran experiments with clone models of varying qubit counts (width). As depicted in Fig. \ref{attack_analysis}c), the 8Q clone model outperformed others, while the 2Q model lagged behind. Hence, an intuitive yet simplistic strategy for an attacker would be to use a high-qubit clone model for training. It is also worth noting that because users are only privy to the output class probabilities, discerning victim models qubit count is virtually impossible.

\begin{table}[!b]
    \vspace{-4mm}
    \centering
    \caption{Quantifying the perturbation added by HVIP and HAVIP compared to generic attack setup for both Top-$1$ and Top-$k$ scenarios on various datasets.}
    \label{perturbation}
    \vspace{-2mm}
    \begin{tabular}{ccccc}
    \cmidrule(lr){2-5}
    \multicolumn{1}{c}{} & \multicolumn{2}{c}{HVIP} & \multicolumn{2}{c}{HAVIP}\\
    \cmidrule(lr){1-5}
    Datasets    & Top-$1$ &Top-$k$ & Top-$1$ & Top-$k$ \\
    \cmidrule(lr){1-5}
    MNIST-4     & 10.29\%      & 6.1\%    &   \textbf{11.2$\boldsymbol{\%}$}   &   \textbf{8.3$\boldsymbol{\%}$}    \\
    Fashion-4     & 8.14\%      & 5.6\%  &    \textbf{15.71$\boldsymbol{\%}$}  &  \textbf{10.2$\boldsymbol{\%}$}            \\ 
    Kuzushiji-4   & 8.43\%      & 4.7\%  &    \textbf{14.29$\boldsymbol{\%}$}   &   \textbf{6.4$\boldsymbol{\%}$}          \\  
    \bottomrule
    \end{tabular}
\end{table}

\noindent \textbf{Mixed vs. Random:} We also assessed the efficacy of a randomly generated dataset against our tailored mixed dataset. In a real-world context, a user of a cloud-based model would typically understand the objective of the QNN—whether it's classification, or another tasks. This understanding would enable them to craft an augmented mixed dataset from related datasets, potentially with analogous images or features. However, if an attacker lacks access to such datasets, their alternative would be to generate a random feature set for querying the victim model. Yet, as observed in Fig. \ref{attack_analysis}d), datasets based on random features might not be a practically viable option since it considerably underperforms compared to the mixed dataset. This is evident in scenarios where the victim returns either the top label (Top-1) or the complete probability vector (Top-\(k\)).

\subsection{Results and Analysis of Defense Strategies}
\noindent \textbf{Defense Configuration:}  For the HVIP evaluation, we primarily trained a cloud-based QNN model on one device for most epochs, then complemented this training on another device for fewer epochs. This ensured our model was resilient to the noise characteristics of both devices. When an attacker sends a query, the victim randomly chooses one of the two devices to execute the model and returns the probability vector. In the HAVIP scenario, we trained two distinct QNN architectures (PQC-1 and PQC-19) on two separate devices. During inference, the victim arbitrarily runs one of the QNNs on its respective trained device. The attacker, utilizing these outputs, inadvertently creates a noisy adversarial dataset $D_A$ which is then used to train clone models, potentially on a distinct device. It's important to note that each device possesses unique characteristics, including varying levels of noise, error rates, and types of basis gates, among other factors.

\begin{figure}[!t]
    \vspace{0mm}
    \centering
    \includegraphics[width=0.9\linewidth]{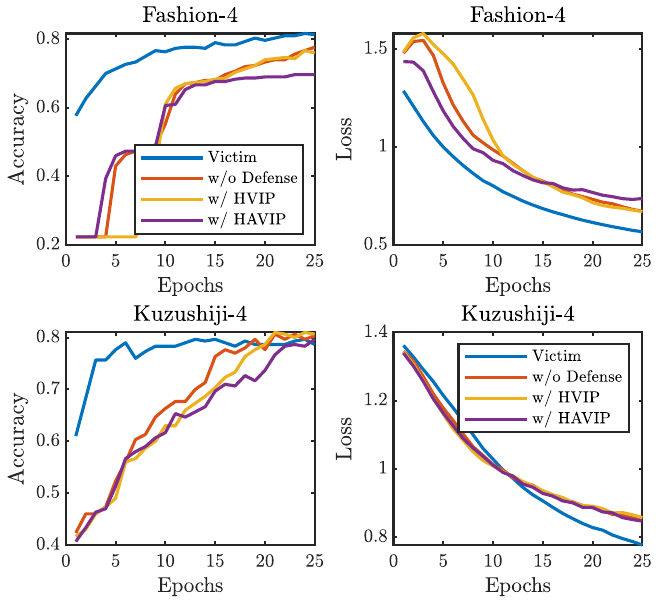}
    \vspace{-4mm}
    \caption{Test accuracy and loss comparison of cloud-based vs. clone models across various datasets, trained using Top-$k$ labels, demonstrating the impact of defense techniques.} 
    \label{defense_plot}
    \vspace{-4mm}
\end{figure}

\noindent \textbf{Perturbation Analysis:} In Table \ref{perturbation}, we evaluate the level of obfuscation introduced by HVIP and HAVIP, compared to a standard attack scenario. This baseline scenario (w/o defense) involves a single cloud-based QNN trained on a specific device. Whenever a query is made, the system always delegates the QNN's execution to this same hardware.
For the Top-$1$ label prediction, we measure the percentage of label mismatches between the defensive strategies (HVIP and HAVIP) and the standard attack scenario. Meanwhile, for Top-$k$ predictions, we assess the total variation distance (TVD) between these defensive approaches and the baseline scenario.
Our findings indicate that merely varying the hardware and architecture introduces perturbations of up to 15.71\%. Increasing the number of devices for querying or increasing the count of trained cloud-based QNNs could further amplify the obfuscation level.

\noindent \textbf{Performance Evaluation:} In Fig. \ref{defense_plot}, we evaluate the efficacy of cloned models across various datasets, both with and without the proposed defenses. The results indicate that cloned models employing the HVIP and HAVIP defenses offer performance either comparable to, or occasionally slightly below, the models without any defenses. For instance, with the Fashion-4 dataset, the clone model's test accuracy declined by approximately $\sim13\%$ with HAVIP, compared to the clone model without defense. Yet, in the other case, the outcomes are nearly indistinguishable across different techniques. One plausible explanation for this observation is that these models are trained in a noisy setting with noisy gradients computed using SPSA and training the clone on noisy datasets potentially enhances its resilience against minor perturbations. This concept aligns with the strategies employed in classical DNNs to prevent adversarial attacks, where models are trained with added noise to the original dataset. This approach not only aids in preventing misclassification but also strengthens the model's resilience to attacks that rely on input perturbations, as the model becomes adept at learning amidst noise. 
Consequently, this renders hardware or architecture-based perturbation methods of limited effectiveness in practice.

\section{Conclusion}
In this study, we explored the effectiveness of model stealing attacks and defenses on QNNs. We observed that such attacks could generate clone QNNs that achieve, on average, $0.95\times$ the test accuracy of the original victim model. Our strategy relied on training substitute QNNs, using the victim as a labeling guide, to form an adversarial dataset. As quantum machine learning evolves towards a cloud-based Quantum MLaaS framework, safeguarding against these attacks becomes paramount. To this end, we evaluate the effectiveness of two defensive methods that capitalize on the intrinsic noise and variability of NISQ devices, aiming to obfuscate the output probability vectors. Although the proposed techniques are capable to perturb the QNN outputs our experimental results indicate that QNN models, when trained on noisy devices, demonstrate resilience to perturbations originating from hardware and quantum architecture variations. As a result, these variations does not impact the efficacy of cloning. Although our initially proposed defense methodologies appeared to have limited effect, this revealed a key insight: QNNs trained in a noisy environment possess an inherent resilience to defenses or attacks based on perturbations/obfuscations. \emph{This study highlights that model theft could be a significant security issue for emerging QML platforms}.

\bibliographystyle{ACM-Reference-Format}
\bibliography{refs}

\end{document}